\documentclass[12pt,preprint]{aastex}

\shorttitle{propagation and dynamics of jets}
\shortauthors{Mizuta et al.}

\begin{document}

\title{Propagation and Dynamics of Relativistic Jets}

\author{Akira Mizuta\altaffilmark{1}, Shoichi Yamada\altaffilmark{2}, 
and Hideaki Takabe\altaffilmark{1}}
\email{amizuta@ile.oksaka-u.ac.jp}

\altaffiltext{1}{Institute of Laser Engineering, Osaka University,
2-6 Yamada-Oka, Suita, Osaka, 565-0871, JAPAN}
\altaffiltext{2}{Department of Physics, Waseda University, Okubo,
Shinjuku, Tokyo, 169-8555, JAPAN}

\begin{abstract}
We investigate the dynamics and morphology of jets propagating into the
interstellar medium using 2D relativistic hydrodynamic simulations.
The calculations are performed assuming axisymmetric geometry
and trace a long propagation of jets.
The jets are assumed to be 'light'
with the density ratio between the beam to the ambient gas much
less than unity.
We examine the mechanism for the appearance of vortices
at the head of jets in the hot spot.
Such vortices are known as a trigger of a deceleration phase
which appears after a short phase in which
one dimensional analysis is dominant.
We find that an oblique shock at the boundary rim near the end of the beam
strongly affects the flow structure in and around
the hot spot.
Weakly shocked gas passed through this oblique shock
and becomes a trigger for the generation of vortices.
We also find the parameter dependence of these effects
for the propagation and dynamics of the jets.
The jet with slower propagation velocity
is weakly pinched, has large vortices
and shows very complex structure at the head of the jets
and extended synchrotron emissivity.
\end{abstract}

\keywords{galaxies: jets --- hydrodynamics -- shock waves 
-- methods: numerical --- relativity}

\section{INTRODUCTION}
It is widely known that there are
three classes of highly collimated and supersonic jets from
dense central objects with accretion disks.
Depending on the central object the following
can be distinguished : protostars,
binary stars, and  Active Galactic Nuclei (AGN).
AGN jets are the largest scale phenomena
and the velocity of the jet beam is highly relativistic,
at least close to
the central objects (see for example, \citet{Pearson81,Biretta99}).
The jet, which originates near an accretion disk
that surrounds an AGN,
can propagate over a long distance up to a few Mpc
while remaining well collimated.
There are two shocks at the end of the jet.
One is a bow shock (or a forward shock) which accelerates the ambient gas.
The other is a terminal Mach shock (or a reverse shock)
at which the beam ends.
At the terminal Mach shock,
non-thermal particles are accelerated and
emit photons due to synchrotron radiation and
inverse Compton scattering.
The gas which crosses the terminal Mach shock
into a hot spot is hottest and pressurized and then
expands laterally and envelops the beam with the shocked ambient gas,
creating a so-called cocoon structure.
At the contact discontinuity between the ambient gas and
the jet in the cocoon
Kelvin-Helmholtz instabilities develop.

Cygnus A is a suitable object to see these features
because it is one of the closest radio galaxies
and the beam propagates to perpendicular direction to
line of sight.
From observations, its size is about 120kpc,
the beam velocity is $0.4 \sim 1c$ and
the hot spot's ram pressure advance speed is $0.03c$ \citep{Carilli98},
where $c$ is the speed of light.
For more details of AGN jets, see for example
\citet{Carilli96,Ferrari98,Livio99}
and references therein.

Other active regions in jets are knots and secondary hot spots.
Some knots with non-thermal emission
can be seen sporadically along the straight beam flow.
The emissivity between the knots varies
even if they belong to the same jet.
Recent high resolution observations of AGN jets
show fine structure of knots in the beam flow
for up to several tens of kilo parsecs
in M87 \citep{Biretta99,Marshall02,Wilson02},
3C273 \citep{Bahcall95,Marshall01,Sambruna02,Jester02},
the jet in Centaurus A \citep{Kraft02},
3C 303 \citep{Kataoka03},
and others \citep{Sambruna01}.
Knots on sub-parsec or parsec scales
are thought to be due to
intermittent flows from the central source
and knots in 'blazars' are due to exceptionally strong shocks
which are caused by
the collisions of internal shocks
\citep{Spada01,Bicknell02}.
Knots on kilo parsec or larger scales are
not well understood.
The high energy particles that are accelerated on parsec scales
may retain their energy for a long time.
In some jets, one observes
secondary hot spots adjacent to the primary hot spot at the head of jets.
The emissivity of secondary hot spot can be as high as that of primary hot spot.
It is thought that the gas in the secondary hot spot is
separating from the primary hot spot.
For example, at both eastern and western side of Cygnus A jet,
a bright secondary hot spot is observed \citep{Wilson00}.
The reason for these multiple hot spots is not well understood.

Recent observations of AGN jets provide information
not only about the features of large scale jets
but also that of very small scale jets
such as in  'Compact Symmetric Objects' (CSOs) which are believed
in the first stage of AGN jets from its morphology.
CSOs show two sided emissivity within the central few kpc
of its parent galaxy.
The physical properties of two hot spots and lobes at both sides 
have been measured \citep{Owsianik98,Polatidis03,Giroletti03}.
CSOs are only a few thousand years old.
and the propagation velocity of the hot spot is a few tens
of percent of the speed of light.
CSOs may tell us 
about the physical conditions during the earliest phases of jets.

To understand the morphology and the dynamics of the jet,
analytical studies and numerical simulations
have been performed for the past thirty years.
\citet{Blandford74} and \citet{Scheuer74}
discussed the structure of the jet with
the relativistic beam model theoretically.
Some of the early numerical work was done by \citet{Norman82}.
Although these were non-relativistic simulations,
they showed the main structure of jets,
that is two shocks at the head of the jet and the cocoon.
The difficulty of numerical relativistic hydrodynamics
has delayed the investigation of the relativistic effects
on the morphology and the dynamics of jets.
Only since the past ten years stable codes with or without
external magnetic fields have been developed for
the ultra-relativistic regime
\citep{Eulderink94,Duncan94,Font94,Koide96,Komissarov99a,Aloy99a,Hughes02}.
which allowed the investigations of
the formation, collimation, and propagation of jets \citep{Rosen99,Koide02} 

\citet{Marti97} and \citet{Aloy99b}
performed long term simulations
to study the morphology of the jets
with 2D and 3D relativistic hydrodynamic codes.
Propagation of the relativistic jets in the external magnetic field
were computed by \citet{Komissarov99b}.
These studies verified that the morphology of jets depends
on a number of dimensionless parameters:
the density ratio of the jet beam to the ambient gas
($\eta\equiv\rho_{b}/\rho_{a}$),
Lorentz factor of the beam,
and the Mach number of the beam ($M_{b}=v_{b}/c_{s,b}$).
Here the subscripts 'b' and 'a' stand for beam and ambient gas,
respectively.
In most of these simulations, a pressure matched jet is assumed,
with the pressure ratio $K\equiv p_{b}/p_{a}=1$.
\citet{Marti98} performed long term simulations of jets
and found a deceleration phase of the propagating jets.
Recently \citet{Scheck02} have pointed out that
the propagation velocity, $v_{j}^{1D}$,
which is derived from one dimensional momentum balance
at the rest frame of the contact discontinuity
is very useful for understanding the propagation
of the jet into the ambient gas.
They assumed a constant velocity
$v_{j}^{1D}=0.2 c$ and the kinetic luminosity of the beam
is fixed to be $L_{kin}=10^{46} \mbox{erg s}^{-1}$ in their models.
Those are the first numerical simulations
using relativistic hydrodynamic code
which assume the propagation efficiency $v_{j}^{1D}/v_{b}$
is less than 0.5
(cf. most numerical simulations done so far asuume the efficiency
is $0.5 \sim 1$).
Thus more wide range of parameter space of
$v_{j}^{1D}$ is necessary
if we agree with recent observations of CSOs are young AGN jets.
\cite{Scheck02} also investigated the dependence of the composition
of jets using a generic equation of state.
They varied the density ratio ($\eta$), pressure ratio ($K$),
Mach number ($M_{b}$) and the fraction of protons, electrons, and positrons
in their three models.
The main difference between the various models
is seen in the radiative properties of the jets.
The deceleration phase is also confirmed and shown to
begin with the generation of a large vortex
at the head of jet.
It seems indeed consistent with observations that
jets have a deceleration phase because CSOs have a propagation
speed of a few tens percent of the speed of light,
while in large scale jets such as
Cygnus A the velocity is only a few percent of speed of light.
However the physics of the generation of vortices is not
well understood.

A first attention to compare the vortex structures
in numerical simulations of the jets with
observational data have been made by \citet{Saxton02,Saxton03}. 
who discussed the shock structures at the head of the jet of 
Pictor A and rings in the cocoon of Hercules A.
We note that radio jets with such features are rare \citep{Gizani03}.

In this paper we will examine the mechanism that causes such vortices in
some detail.
We present three simulations with different $v_j^{1D}$
as injected beam conditions.
We pay special attention to the generation of vortices 
in the hot spot and the separation of vortices from hot spots
which strongly affects the dynamics of jets.
We show the mechanism of generation of vortices at the head of jet.
We hope that this study will help us to better understand
the propagation of jets into the ambient gas.
This paper is organized as follows.
In Sect. 2, the basic equations are presented.
In Sect. 3, we show our three models for the numerical numerical simulations.
The results and discussion are shown in sect. 4.
The surface brightness of synchrotron emissivity and the resolution
study are also shown.
The conclusion is given in Sect. 5.

\section{BASIC EQUATIONS}
\subsection{RELATIVISTIC HYDRODYNAMIC EQUATIONS}
Relativistic flows, $v \sim 1$, affect
the morphology and dynamics of jets.
The relativistic hydrodynamic equations \citep{Landau}
then have to be solved:
\begin{eqnarray}
\label{renzoku}
{\partial (\rho W) \over \partial t}+
{1\over r}
{\partial r(\rho W v_{r}) \over \partial r}+
{\partial (\rho W v_{z}) \over \partial z}=0,\\
\label{momen}
{\partial (\rho hW^2 v_{r}) \over \partial t}+
{1\over r}
{\partial r(\rho hW v_{r}^2+p) \over \partial r}+
{\partial (\rho hW^2 v_{r}v_{z}) \over \partial z}={p\over r},\\
{\partial (\rho hW^2 v_{z}) \over \partial t}+
{1\over r}
{\partial r(\rho hW^2 v_{r}v_{z}) \over \partial r}+
{\partial (\rho hW^2 v_{z}^2+p) \over \partial z}=0,\\
\label{energy}
{\partial (\rho hW^2 -p) \over \partial t}+
{1\over r}
{\partial r(\rho hW^2 v_{r}) \over \partial r}+
{\partial (\rho hW^2 v_{z}) \over \partial z}=0,
\end{eqnarray}
where, $\rho, p, v_{i}, W,$ and $h$
are rest mass density, pressure, three velocity component in
$i$ direction, Lorentz factor ($\equiv (1-v^2)^{-1/2}$),
and specific enthalpy 
($\equiv 1+\epsilon +p/\rho$), respectively.
In this section we equate the velocity of light to unity.
We assume axisymmetry.
The set of equations (\ref{renzoku}-\ref{energy}) is not closed
until the equation of state is given.
In this paper we assume an ideal gas;
\begin{eqnarray}
\label{EOS}
p=(\gamma -1)\rho\epsilon ,
\end{eqnarray}
where $\gamma$ and $\epsilon$ are a constant specific heat ratio
and specific internal energy, respectively.

\subsection{PROPAGATION VELOCITY}
\label{1dmoment}
A one dimensional analysis is useful for estimating
the propagation velocity of the jet.
Here the propagation velocity $v_{j}$ is defined to be
the velocity of the contact discontinuity at the end
of the jet in the rest frame of the ambient gas.
For very 'light' jets ($\eta \ll 1$),
in general, the propagation velocity is less
than that of the beam because the conversion
of the large fraction of the beam kinetic energy
to the thermal energy occurs
at the terminal Mach shock.

\subsubsection{NON RELATIVISTIC CASE}
If the velocity and temperature of the gas
are in the non-relativistic regime,
it is sufficient to consider the Euler equation.
Assuming momentum balance in the rest frame
of contact discontinuity,
\begin{eqnarray}
\label{nonmomentum}
S_b \left( \rho_{b}(v_{b}-{v_{j}}^{1D})^2+p_{b} \right)=
S_a \left( \rho_{a}(-{v_{j}}^{1D})^2+p_{a} \right),
\end{eqnarray}
where $S_b$ and $S_a$ are the cross section of the beam and the hot spot,
and ${v_{j}}^{1D}$ is the propagation velocity derived from
one dimensional analysis.
Solving Eq.(\ref{nonmomentum}) for ${v_{j}}^{1D}$, we obtain
\begin{eqnarray}
{v_{j}}^{1D}=
{A \eta v_{b}-\sqrt {A^2\eta^2 v_{b}^2-(A\eta -1)
(A\eta v_{b}^2+(AK-1)c_{s,a}^2/\gamma)}\over
A \eta -1},
\end{eqnarray}
where $c_{s}$ is sound speed and $A$ is cross section ratio ($A\equiv S_b/S_a$).
If we can assume $A=1$ and
neglect the term that includes $K(\equiv p_{b}/p_{a})$ in
Eq.(\ref{nonmomentum}),
then this equation reduces to
\begin{eqnarray}
{v_{j}}^{1D}=
{\sqrt {\eta}\over \sqrt{\eta} +1} v_{b}.
\end{eqnarray}
This is the same as the equation derived by \citet{Norman83}
for the pressure matched jet ($K=1$).
If the density ratio $\eta$ is much less than unity,
in the case of 'light' jet,
the propagation efficiency defined as $v_{j}/v_{b}$
is much less than unity.

\subsubsection{RELATIVISTIC CASE}
For the relativistic case,
assuming again 1D momentum balance
between the beam and the ambient gas
in the rest fame of the contact discontinuity at the head of the jet,
\begin{eqnarray}
S_b \left( \rho_{b}h_{b}W_{j}^2W_{b}^{2}(v_{b}-v_{j}^{1D})^2+p_{b} \right)
=S_a \left( \rho_{a}h_{a}W_{j}^2(-v_{j}^{1D})^2+p_{a} \right),
\end{eqnarray}
where $W_{j}=(1-{v_{j}^{1D}}^2)^{-1/2}$.
This equation leads to
\begin{eqnarray}
\label{1Dvelo1}
 v_{j}^{1D}
={A\eta _{R}^{*}v_{b}-
  \sqrt{A^2{\eta _{R}^{*}}^2v_{b}^2
-(A\eta ^{*}_{R}-1)(A\eta _{R}^{*}v_b^2+(AK-1){c_s}_a^2/(\gamma W_j^2))}
 \over A\eta ^{*}_{R}-1},\\
\mbox{with}\ \ \eta_{R}^{*}\equiv \eta_{R}W_{b}^2,
\ \ \eta _{R}\equiv {\rho_{b}h_{b}\over \rho_{a}h_{a}}.
\end{eqnarray}
In general, the sound speed of interstellar matter
is much smaller than the velocity of the beam
$(c_{s,a}\ll v_{b})$¡¤
so we can neglect the term including $K$.
As a result, Eq.(\ref{1Dvelo1}) becomes
\begin{eqnarray}
\label{momentjet}
\displaystyle{
v_{j}^{1D}={\sqrt{A\eta_{R}^{*}}\over 1+\sqrt{A\eta_{R}^{*}}}v_{b}}.
\end{eqnarray}
This is equal to the equation
derived by \citet{Marti97} for the pressure matched jet
($K=1$) if we assume $A=1$.
\citet{Scheck02} showed that the time evolution of
the actual propagation velocity is
almost the same in their models for fixed $v_{j}^{1D}$ and $A=1$,
in spite of the large variation of the specific internal energy of the
beam during the first phase.
There are several ways to explain the deceleration of jets
between sub parsec and mega parsec scales.
If the density ratio $\eta$ varies, $\eta$ should become smaller than that
in the earlier phase.
The Lorentz factor should also become smaller.
The velocity at the head of the beam should be relativistic
to reproduce strong emissivity at the hot spot.
The effect of the Lorentz factor seems small.
Another possibility is variation of the density profile of the ambient gas. 
For example, it is natural to assume that the ambient gas
decreases in density as function of distance from the AGN.
In the models of \citet{Scheck02}, the multidimensional effect,
namely decreasing cross section's ratio, becomes very
important for the dynamics and the jet decelerates gradually.
\citet{Scheck02} compared the time evolution of the dynamics
with the extended Begelman-Cioffi model.
The original model derived by \citet{Begelman89} assumes
a constant propagation velocity and
\citet{Scheck02} extended the assumption to a power law.

\section{NUMERICAL CONDITIONS}
We solve the relativistic hydrodynamic equations
Eq.(\ref{renzoku}-\ref{EOS}) with a 2D relativistic
hydrodynamic code recently developed by one of the authors (A.M.).
The detailed numerical method and test calculations of the code
are given in the appendix of this paper.

We use a 2D cylindrical computational region
$r \times z$.
The grid size in both $r$ and $z$ directions is
uniform, namely $\Delta r=\Delta z=const.$
We assume that the ambient gas is homogeneous initially.
Calculations including clouds in the ambient gas
have been made by \citet{Pino99,Zhang99,Hughes02}.
A relativistic beam flow $(v_{b})$ which is parallel to $z$ axis
is continuously injected from one side
of the computational region ($z=0$).
The inner 10 grid points from the symmetry axis are used for this.
The radius of the injected beam $R_{b}=10\Delta r$ is used
as a scaling unit in this study.
The computational region covers
$50 \mbox{R}_{\mbox{b}} (r) \times 180 \mbox{R}_{\mbox{b}} (z)$;
corresponding to a grid $500\times 1800$ grid.
The radius of the beam near the central engine is
unknown,
and we assume a plausible value of $1\mbox{R}_{\mbox{b}}=0.5\mbox{kpc}$,
corresponding to $25\mbox{kpc}\times 90\mbox{kpc}$
for the computational region.
A free outflow condition is employed
at the outer boundaries for $r=50 \mbox{R}_{\mbox{b}}$ and
$z=180 \mbox{R}_{\mbox{b}}$.
A reflection boundary is imposed on the symmetry axis as well as at $z=0$
with $r >\mbox{R}_{\mbox{b}}$.
The boundary condition at $z=0$ is crucial for dynamics and
the outer shape of the jets.
The free boundary condition permits the gas to escape at the back side
\citep{Saxton02}.
According to observations jets have counter jets which propagate
in the opposite direction.
We assume the reflective boundary condition so that
our calculations begin near central engine.

We examine three injection beam conditions (Table \ref{condition})
with different ${v_{j}}^{1D}$.
The description of ${v_{j}}^{1D}$ is given in Eq.(\ref{momentjet}),
here $A=1$ is assumed.
The beam velocity is fixed to $v_{b}=0.99 c$
(the corresponding Lorentz factor is $W_{b}=7.1$).
The Mach number of the beam is also fixed to $M_{b}=6.0$.
We assume that the beams don't have highly relativistic temperatures
and that remains constant $v_{b}$.
As a result ${v_{j}}^{1D}$ depends only on the density ratio $\eta$.
\citet{Scheck02} adopted ${v_{j}}^{1D}=0.2 c$
as the injected beam conditions for their calculations.
This is in the range of
acceptable propagation velocity based on observation of CSOs.
We explore the region of the propagation velocity ${v_{j}}^{1D}$
from $0.2 c$ to $0.4 c$ and investigate the dependence of
the difference of ${v_{j}}^{1D}$ caused by different $\eta$.
With increasing ${v_{j}}^{1D}$,
$\eta$ varies about a factor of ten; from $1.28\times10^{-3}$ to $9.15\times10^{-3}$.
We label theses cases JB02, JB03, and JB04,
and the corresponding ${v_{j}}^{1D}$ are $0.2 c$, $0.3 c$, and $0.4 c$.
Although there is only about one order difference in
the density ratio for all beams,
doubling value of ${v_{j}}^{1D}$
has a big effect on the dynamics of jets and their long term propagation.
Because the spatial scale is fixed,
we adopt different expiration times for each models,
see Table \ref{condition} again.

The initial rest mass density of ambient gas $\rho_{a}$
is unity for all models.
The pressure ratio($K$) is chosen
to be from $K=10$ to $100$
so that the ambient gas is approximately similar in all cases.
As a result, the temperature of the ambient gas for all injected beams
is a few eV, if we assume the ambient matter is pure hydrogen gas.

Because we choose similar values of $v_{j}^{1D}$
to those of \cite{Scheck02},
other parameters ($\eta, K, M_{b}$) are also very similar.

\section{RESULTS AND DISCUSSION}
\subsection{MORPHOLOGY AND DYNAMICS}
From our simulations,
we find that the overall morphology and dynamics of the jets
are similar to that discussed in previous work.
Figures  \ref{contourA}, \ref{contourB} and \ref{contourC}
show snapshots of rest mass density,
pressure, and Lorentz factor for
three models near the end of the simulations,
$t=1770[\mbox{R}_{\mbox{b}}/(c)]$(model JB02),
$t=1140[\mbox{R}_{\mbox{b}}/(c)]$(model JB03) and
$t=570[\mbox{R}_{\mbox{b}}/(c)]$(model JB04).
Since these jets have different beam conditions,
the end times are also different as discussed in the previous section.
However this is not only due to different ${{v_{j}}^{1D}}$ values,
but also because of differences in deceleration phase of the jets
which are caused by the generation and separation of vortices
at the head of the jet, as we will discuss later.
At first glance, the outer shapes of the jets are quite different.
Model JB02 shows a conical outer shape which is  very similar to the results
found by \citet{Scheck02}.
This is not seen in the other two models (JB04 and JB03).
All beams remain collimated from $z=0$, where the beam is
injected into the computational region, to the head of the jets.
It is also important to note that the beam radius dose not increase
monotonically from the source to the head of jets.
At the head of the jet the radius of the beam is $3\sim 5 R_{b}$.
The opening angle is very small $1^{\circ}\sim 2 ^{\circ}$ for all models.
High Lorentz factors exist only along the $z$ axis in the beam flow,
but they do vary.
That means that a large part of the injected kinetic energy is
transported from the central engine to the head of the jet
along the beam flow.
At the end of the beam a strong 'terminal Mach shock'
can be seen.
One of the most active points is called a 'hot spot' into which
shocked beam gas enters through the terminal Mach shock
at the head of the jet.
The pressure in the hot spot is very high due to the
energy dissipation at the terminal Mach shock
and is matched by ambient gas compressed at the bow shock.
Moderate Lorentz factors ($W\sim 2$) exist in the cocoon.
These are due to back flows
that begin at a hot spot and flow back in the center of the cocoon
parallel to the beam flow.
This mildly relativistic back flow seems a little strange thing 
because the head is proceeding very slow ($\lesssim 0.2$) and
the expanding velocity from the hot spot
is up to 0.5c in the comoving frame (maximum sound speed).
It should be noted that this mildly relativistic backflow
is longer for slower jets.
Thus some acceleration mechanism or other effects are necessary.
We discuss this in the next subsection.
This back flow creates a shear flow
and the contact surface becomes unstable
due to Kelvin-Helmholtz instabilities.
The surface between the back flow and
shocked ambient gas's flow also becomes unstable
and causes the appearance of vortices in the larger cocoon.

\begin{figure}
\plotone{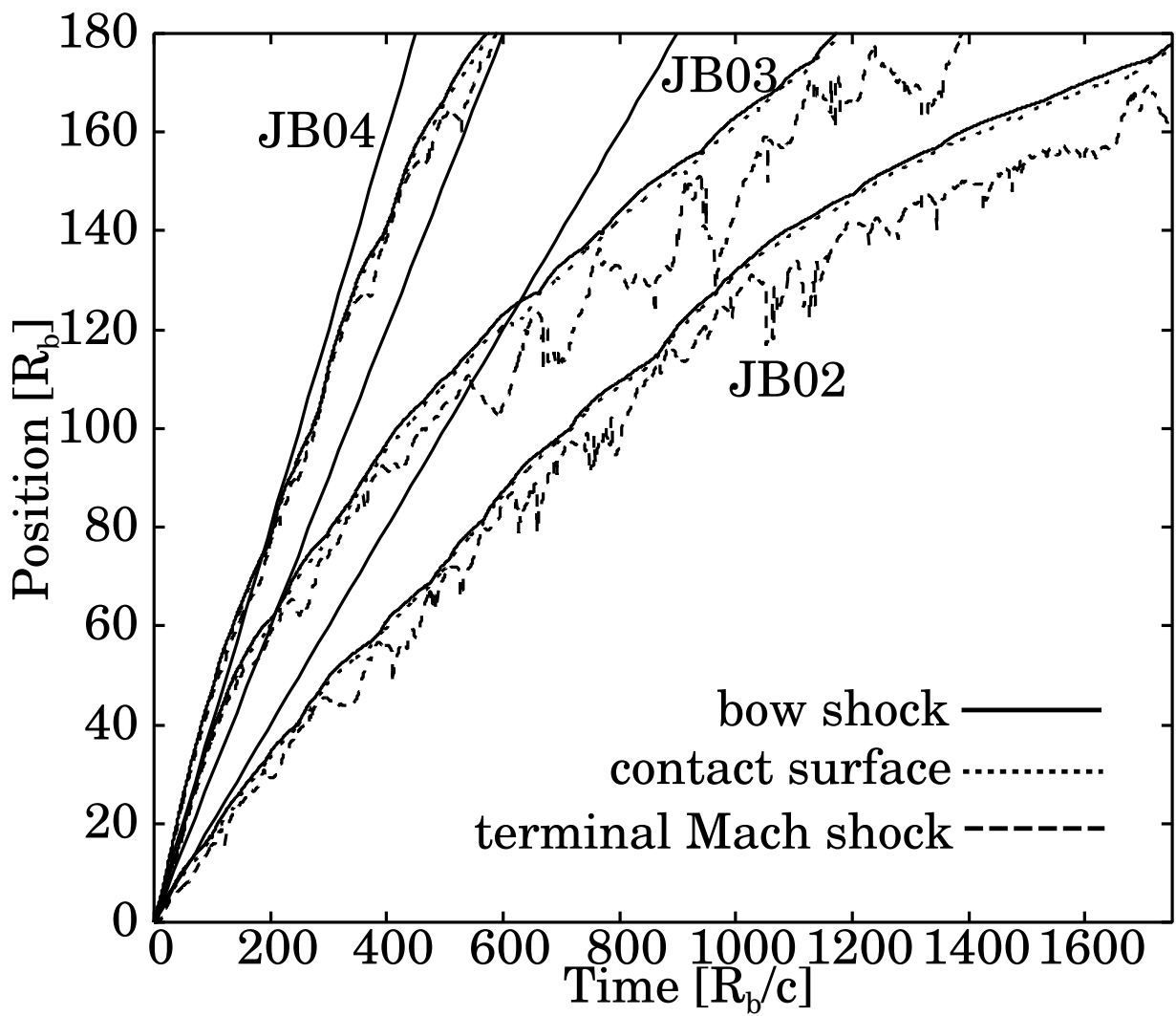}
\caption{Time evolution of the position of
the bow shock, the contact discontinuity, and the terminal Mach shock
for models JB02, JB03 and JB04.
\label{1Dsurface}}
\end{figure}

To understand the difference of these models after long term propagation
we must study the dynamics of the jets.
Figure \ref{1Dsurface} shows
the time evolution of the positions of
the contact discontinuity, the bow shock (forward shock), and
the terminal Mach shock (reverse shock)
at the head of the jet for each model.
Three straight lines which correspond to
the lines of one dimensional analysis are also shown
for the comparison.
It is difficult to define the positions of these surfaces
because of the complex structure at the head of the jets.
In this paper we show the positions at $r=0$, namely along the $z$ axis.
It is hard to identify the contact discontinuity between
shocked ambient gas and shocked beam gas especially in later phase
because of mixing of the shocked beam with the ambient gas due to the
generation of vortices.
The contact discontinuity is defined as the boundary of shocked ambient gas
that has half the maximum density.
We plot the points where the Lorentz factor becomes 2 at the head of the
jet.
The two phases indicated by \citet{Marti98} and \citet{Scheck02}
are also seen in our results.
During the first phase, all of the slopes are constant.
Model JB02 has a propagation velocity $\sim 0.2c$
and the one dimensional analysis discussed in section
\ref{1dmoment} is in good agreement with our result.
For JB03 and JB04 the propagation velocity is a little faster
than that of in our one dimensional estimates.
During first phase the surfaces are very close each other.
On the other hand, in the second phase,
the surfaces separate and approach each other repeatedly.
The propagation velocity is no longer constant and
the jet is decelerating gradually.
Some vortices grow at the head of jet and separate towards the back side.
This affects the dynamics of the jets.

Figure \ref{velocity} shows snap shots of
the velocity ($\sqrt {v_{r}^2+v_{z}^2}$)
during the earlier phase of the each models.
Although its path is not as straight as that of the beam,
the back flow does remain parallel to the beam flow.
The back flow velocities are very similar for all models
($0.3\sim 0.4c$).
Because the propagation velocity is less than
that of back flow for the model of JB02,
most of it reaches the boundary,
after which some of the gas expands in lateral direction,
and other gas form a new flow which is between the beam and
the back flow.
This new flow also can be seen in the case of JB03 near the boundary,
but it does not reach the head of jet like JB02 because
the propagation velocity is faster than that of JB02.
The beam flow and back flow lie side by side in the model
JB03 and JB04 near the head of the jets.

In later phases this 'third' flow, which is not beam flow but
has the same propagating direction, exists in all models
because of vortices generated in the cocoon
due to Kelvin-Helmholtz instabilities.
At this stage the back flow becomes very complex.
How much gas reaches the boundary at $r=0$
affects the expansion into lateral directions
for each model.
Slower propagation velocities result in cone like outer shapes of the jets.
The shocked ambient gas forms
a shell-like structure like a football,
resembling the observed X-ray emission Cygnus A by
{\it Chandra} \citep{Wilson00}.
It is consistent with our results
because the present propagation velocity of Cygnus A is very small
($\sim 0.03c$).
A long time may have passed after the deceleration phase began in the Cygnus A jet.
In contrast, in models JB03 and JB04, which have
propagation velocities faster than that of the back flow in the earlier
phase
do not have a conical shape.
These jets are decelerating.
The effect of a decreasing propagation velocity can be seen
near the head of the jet in JB03,
showing a conical shape around the head of the jet.

\subsection{VORTEX FORMATION IN HOT SPOTS}
We discussed the generation and separation of large vortices at the head of the
jet which occurs repeatedly during the second phase.
Such processes strongly affect the dynamics of jets and was
also found by \citet{Scheck02}.
It is also consistent with observations
between the high speed CSO sources and slower large scale jets.
The reason for this deceleration seems to be the formation of vortices.
Where and how are these vortices created ?
The possibilities of hydrodynamic instabilities at the head of jets
have been discussed before.
\citet{Norman82} discussed that there are
Rayleigh-Taylor instabilities at the contact discontinuity.
Recently, \citet{Krause03} mentioned the Kelvin-Helmholtz instability
between the flow from hot spot and shocked ambient gas.

To understand the generation of these vortices
we need to focus on the flow structure in and around the hot spot.
The vortices appear in the hot spot
and they originate by the flow through an oblique shock at the end of the beam.

\begin{figure}
\begin{center}
\resizebox{12.0cm}{!}{\plotone{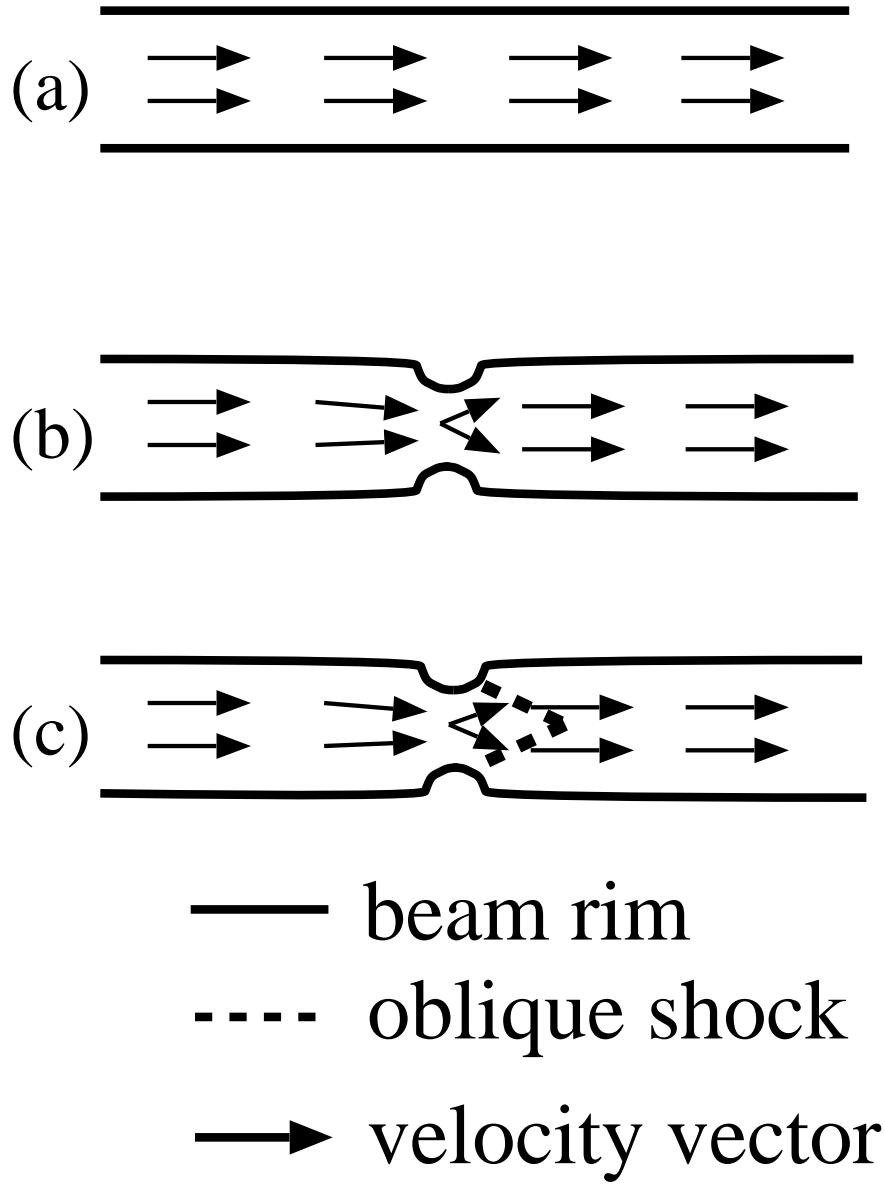}}
\caption{A schematic of appearance of an oblique shock in the beam
(not scaled)
\label{obshock}}
\end{center}
\end{figure}

It is known that oblique shocks appear within the beam,
even before the end,
when the beam expands in lateral directions.
This re-establishes pressure equilibrium between the beam
and its surroundings and keeps it confined (Fig \ref{obshock}a).
When the beam is pinched for some reason,
the gas tends to expand due to increased pressure caused by the compression
(Fig \ref{obshock}b).
If the pressure outside is high enough to confine the beam,
an oblique shock appears to prevent it from expansion.
(Fig \ref{obshock}c).
The radius of the beam after reconfinement
depends on the pressure outside of the beam
and how the beam is pinched.

\begin{figure}
\begin{center}
\resizebox{11.5cm}{!}{\plotone{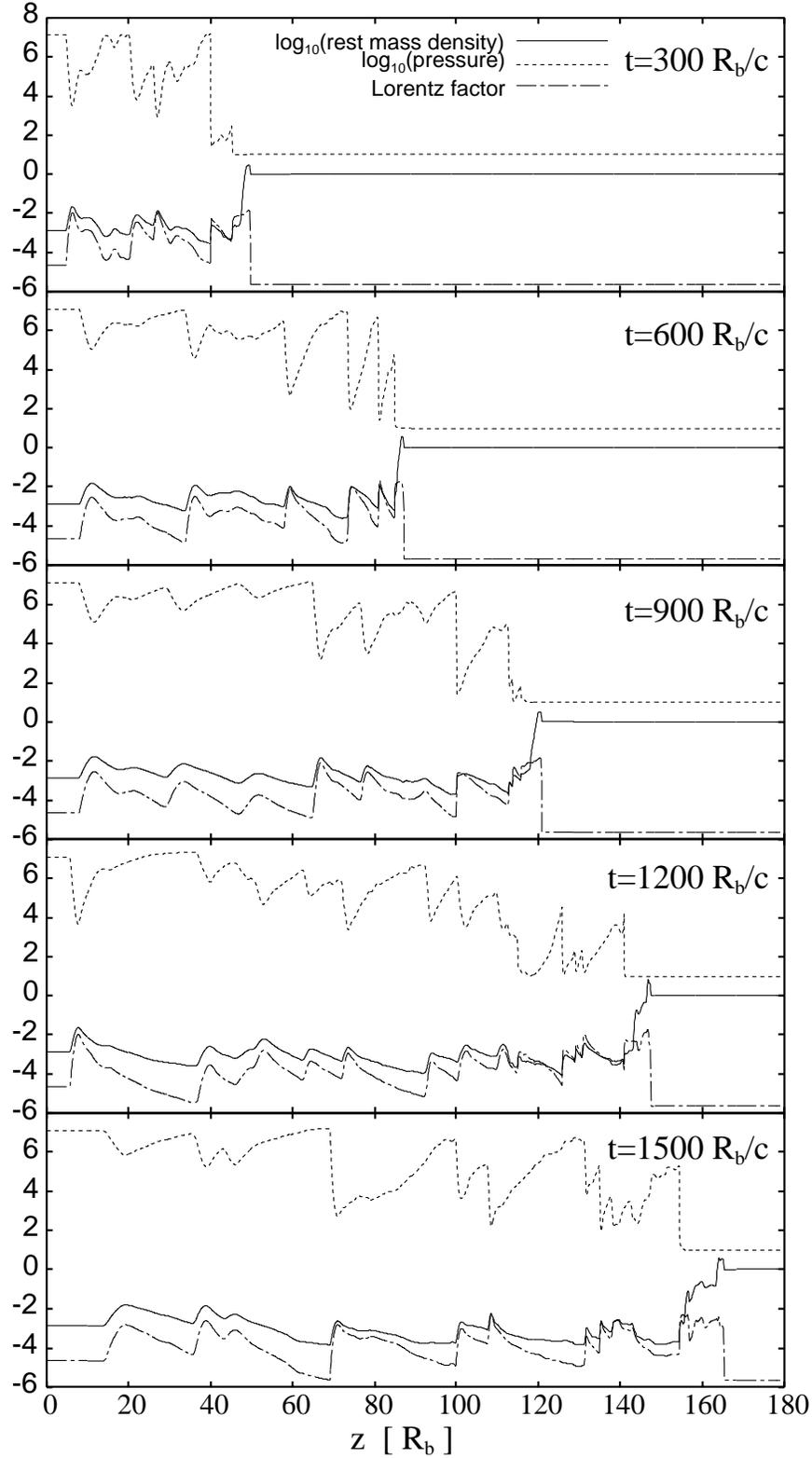}}
\caption{The log scale rest mass density (solid), log scale pressure (dashed),
 and Lorentz factor (dashed point) profile
along the z axis of model JB02 at t=
$300,600,900,1200, \mbox{and} 1500 {\mbox{R}}_{\mbox{b}}/(c)$.
\label{1dplot}}
\end{center}
\end{figure}

Figure \ref{1dplot} shows the rest mass density, pressure and Lorentz factor profile
along the $z$ axis of model JB02
at $t=300,600,900,1200,$ and $1500 {\mbox{R}}_{\mbox{b}}/c$.
The beam is confined by the pressure of
the cocoon and shocked ambient gas.
However the surface of the beam is not stable.
In our calculations the first oblique shock appears where the beam
is injected.
Such shocks appear irregularly in the beam.
As these shocks have different speeds,
a slower shock is caught up by a faster shock.
This might explain the high emissivity knots in the jets.
When an oblique shock appears at the end of the beam,
most of the gas passes a terminal Mach shock with large
dissipation.
However, some of the gas passes through the
oblique shocks on the side (Fig. \ref{vortex} (a)-(b)).
If the angle is small, the oblique shock is very weak.
The loss of kinetic energy should be small but the pressure becomes
as high as that in the cocoon.
Then a mildly relativistic back flow begins.
The effect of these oblique shocks can be seen
in slower jets.
This is why the mildly relativistic flow
is longer for slower jets.
Such a fast velocity flow through the oblique shock
can propagate further in lateral direction
than slower gas from the hot spot before it propagates backwards.
The flow path becomes then a circular, arc like vortex.
This vortex can sometimes, which triggers instabilities at
the beam surface and internal oblique shocks, reach the beam flow.
The surface of the beam becomes more unstable and has oblique shocks
there.
The oblique shocked flow has an important effect on the gas in the hot spot
since it blocks further outflow.
In the hot spot the velocity is not constant.
The nearer the gas is to a corner that is a crossing point
between a contact discontinuity and $z$ axis,
the slower the velocity.
Although the outflow from the hot spot is blocked,
the beam flow continues to enter the hot spot.
This then drives a clockwise rotation vortex
in the hot spot (Fig. \ref{vortex} (c)-(e)).
This vortex is very important for the dynamics
because the increasing radius of vortices causes an increase of cross
section of the beam.
As a result, the beam decelerates (see Eq.(\ref{momentjet})).
The vortex can grow till the radius becomes about twice large as
that of the beam.
The slower a jet is, the larger vortex it can support
because the shocked ambient gas which is just going to backward
has smaller momentum to this vortex.
When a vortex grows, the jet decelerates.
On the other hand when a vortex separates (Fig. \ref{vortex}
(f)-(g)),
the jet accelerates slightly.
But the next vortex grows soon.
In Fig. \ref{1Dsurface} the size of hot spot, namely, the distance between
a contact discontinuity and a terminal Mach shock,
is oscillating with increasing amplitude.
This occurs in all of models but
is strongest in the slower jets. 
The increase of the amplitude of the oscillation
increases the growth time of a vortex.
A long growth time causes a larger radius of the vortex.
The dynamics is a repetition of these processes
during the second phase.

Most of gas in this vortex has passed trough the 'terminal Mach shock'.
The vortex also contains non thermal particles accelerated by this shock.
A separating vortex can be observed as an active region around the hot spot.
The best candidate for this region is the secondary hot spot
discussed in the introduction.
The high energy particles loose their energy quickly,
so that a separating vortex is a little less bright than
that of primary hot spot.

When a jet accelerates, the bow shock has a nose cone like shape.
On the other hand, the bow shock which is decelerating has a flat shape.
This may be observed in the eastern lobe of Cygnus A,
where the hot spot seems to be slightly ahead of the lobe.
This may be because the vortex may just have separated and
the jet is in a brief accelerating phase now.

Once a jet exhibits such oblique shocks and vortices,
the flow structure around the hot spot becomes more complex.
The effect of this turbulence is strong in the case of slowly
propagating jets
because the confinement effect of shocked ambient gas is weak.
The beam is pinched when a circular arc like flow from the oblique shock
reaches the beam or a separated vortex moves backwards.
The end of the beam tends to expand and has more oblique shocks there.
The beam is sometimes constricted in the middle by such effects
and a strong shock appears.
This shock may be related to knots observed in the beam.

During the later phase of the model JB02,
the size of hot spot becomes very large and clumpy.
As a result, the gas which passed the terminal Mach shock
follows the edge of this clumpy gas.
Some of it is mixed in at locations where the gas
reaches the shocked ambient gas and back flow begins.
This then also separates from the head of the jet and
'normal' hot spot appears again.
The large clumpy gas does not display a bright 'hot spot'.
It corresponds to a 'temporary absent' hot spot 
discussed by \citet{Saxton03} and
can be a good candidate to the absent hot spot in Hercules A.

\subsection{SYNCHROTRON EMISSIVITY}
Recent some numerical simulations of jets show the surface brightness
of the synchrotron emissivity \citep{Scheck02,Saxton02,Saxton03,Aloy03}.
We follow the analysis by \citet{Saxton02,Saxton03}
and compare our results with theirs and observations.

We use an approximation which assumes the synchrotron emissivity
is proportional to $pB^{1+\alpha}$, where $p$ is pressure,
$B$ is magnetic filed, and $\alpha$ is the spectral index.
We use $\alpha =0.6$ that is typical value.
The magnetic pressure $\sim B^2$ is assumed in equilibrium with
thermal pressure $p$.
The surface brightness is derived from revolved 2D numerical results.
The emissivity is $fp^{1.8}$, where $f$ is the fraction of the gas
which originats the beam.
The advection equation of $f$ is transformed to conservative form
using mass conservation equation and
solved with hydrodynamic equations.

Figure \ref{emissionJB02} and \ref{emissionJB03}
show the emissivity of the models JB02 and JB04
in log scale with four order magnitude from maximum intensity
using gray scaled color bar (white is the brightest emissivity)
in each panel is shown.
Different several angles,
$\theta=0,15,30,45,60,75,$ and $90 ^{\circ}$, where $\theta$ is the angle
between $z$ axis and the line of sight
are assumed.
JB02 has very extended emissivity from the head to
the root of the jet.
On the contrary,
JB04 which indicates less complex structure
has the emissivity only around the head.
Thus, the appearance of vortices causes
the extension of emissive regions toward backside. 
The brightest region corresponds to the hot spot.
The other bright regions which corresponds to
the oblique shocks in the beam appear irregularly.
This may correspond to the observed knots in the beam.
The same emissivity is usually seen in the observations of the jets.
Both of models have a ring like emissivity near the head of the jet
when the angle $\theta$ is $30 \sim 75^{\circ}$.
This feature is also shown and discussed the similarity with
observation of Hercules A in \citet{Saxton02,Saxton03}.
Several rings in the lobe has different radius and emissivity.
The observations of Hercules A by \citet{Gizani03}
resemble our results of emissivity in JB02.
These results indicate that the lobe of Hercules A has
a complex structure with vortices.

\subsection{RESOLUTION DEPENDENCE}
At last, we discuss the dependence of the resolution for
our discussion.
The calculations with higher resolution allow
to have finer structures
because how much numerical dissipation is included
depends on the resolution.
Especially for the problem which has the appearance of
vortices, such an effect is very important
and may affect the size of vortices and the timing of
formations of vortices.
We performed calculations with 1.5 times and twice higher resolution
for our three cases discussed in the previous sections,
but the scales are restricted to save CPU time.
In other words 15 (1.5 times resolution)
and 20 (twice resolution) grid points are used for innerlet beam.

\begin{figure}
\begin{center}
\resizebox{13.0cm}{!}{\plotone{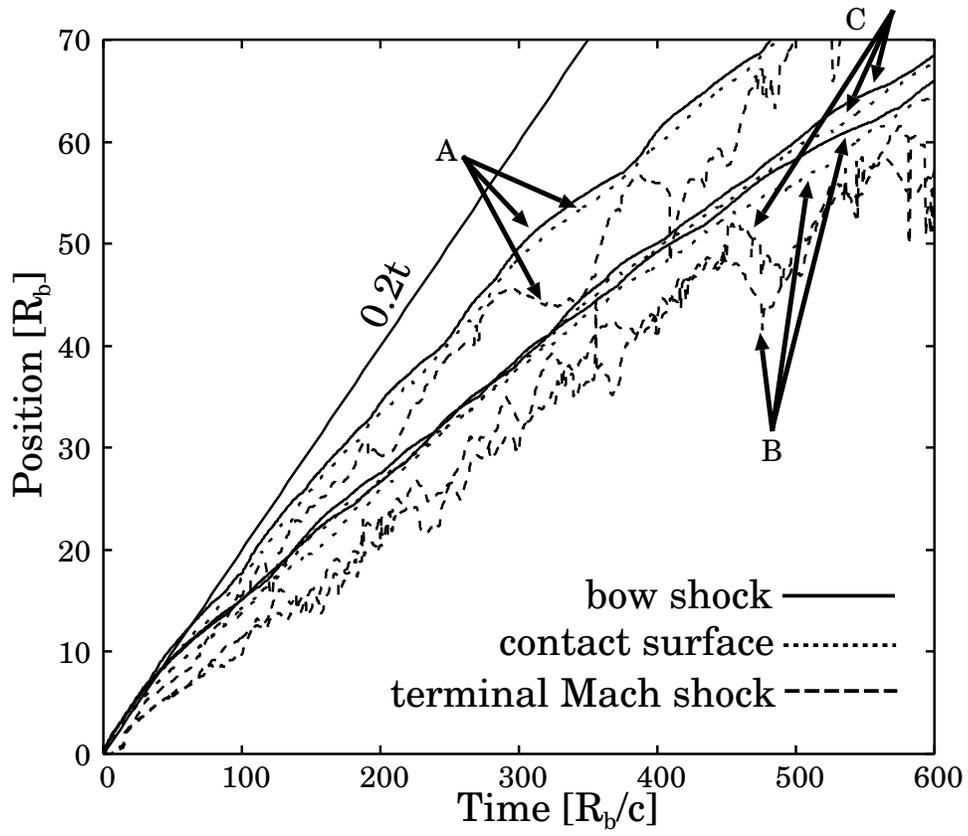}}
\caption{Time evolution of the surfaces
at the head of the jets (same as \ref{1Dsurface} but only JB02)
using different resolution for the calculation.
A (innerlet beam 10:normal case), B (15), C(20).
\label{highreso}}
\end{center}
\end{figure}

Higher resolution calculations also show the same property
as discussed in the previous subsections,
namely, two phases are observed,
the jets have the generation and separation vortices
at the head of the jet
and these effects are strongly seen in the slower injected beam models.
Figure \ref{highreso} shows the time evolution of the positions
of the bow shocks, contact surfaces, and terminal Mach shocks
with different resolutions for the condition of JB02.
This figure is as same as Fig. \ref{1Dsurface},
but only JB02 case is shown and the scales are restricted.
The jets with higher resolution propagate a little slower
than that of normal version.
This is caused by a large separation of a terminal Mach disk
from the contact surface in the early second phase
because the large separation allows the appearance of large
hot spot which is seen in the later phase of the calculation
in JB02 with normal resolution and large cross section
for the efficiency of the propagation. 

\section{CONCLUSION}
Three numerical simulations of relativistic jets are shown in this paper.
We pay special attention to the formation of vortices and its
parameter dependence.
The propagation velocity derived from
one dimensional momentum balance is varied from $0.2c$ to $0.4c$.
These estimations for the velocity are based on recent observations of CSOs.
We identify two phases based long term simulations,
confirming previous results.
The propagation velocity during the first phase
may be estimated using a one dimensional analysis.
The length between the bow shock and the contact discontinuity does
not change significantly.
On the other hand, the distance between the terminal Mach shock
and the contact discontinuity oscillates and
its amplitude increases gradually
during the second phase.
This effect is strongest in slow jets.
During the second phase,
an oblique shock with a small angle at the end of the beam
appears.
This flow dramatically affects the dynamics and morphlogy,
and emissivity of the jets in the second phase.
Weakly shocked gas then foams a back flow without
large energy dissipation.
The velocity of this flow is faster than that of expanding flow
from the hot spot.
The hot spot is blocked by this weakly shocked flow.
As a result, a vortex forms in the hot spot. 
When a vortex grows, the increasing cross section decelerates the jet.
When a vortex separates from the jet,
the jet accelerates.
After this a new vortex appears and
this process occurs repeatedly during the second phase.
The jet decelerates gradually in time.
As a future work,
the effect of the generation and separation of vortices at the head of
the jet
should be investigated using three dimensional numerical simulations.

The synchrotron emissivity of the jets are shown
The strong emissivity of the synchrotron radiation at the hot spot
appear.
The irregular emissivity on the $z$ axis appears and
had same similarity with observed knots in the beam.
In the slowest beam model
very extended emissivity is shown from the head to the root of the jet and
the ring like structure of unusual emissivity of Hercules A
is reproduced.
In the fastest beam model
the emissivity can be seen only around the head of the jet
and along the beam.
The separation of vortices from the head of the jet
strongly affects the extension of the emissivity.

\

This work was carried out on NEC SX5, Cybermedia Center and
Institute of Laser Engineering, Osaka University.
We appreciate computational administrators for technical supports.

A.M. acknowledges support from the Japan Society for the Promotion of
Science (JSPS).
A.M. would like to thank N. Ohnishi, H. Nagatomo, and
K. Sawada for useful suggestions to the numerical methods.
We also gratefully acknowledge
J. M$^{\underline{\mbox{a}}}$. {Ib{\' a}{\~ n}ez}, S. Koide,
M. Kino, T. Yamasaki for helpful discussions on AGN jets.
We acknowledge comments on the emissivity by C. Saxton and G. Bicknell.
We appreciate useful comments and suggestions by
W. van Breugel and anonymous referee
which have improves this manuscript.

\begin{appendix}
\section{RELATIVISTIC HYDRODYNAMIC CODE}
In this appendix, we describe a numerical method used in
our relativistic hydrodynamic code and show some results of
typical test problems for relativistic hydrodynamic codes,
namely, the shock tube problem and strong reflection shock problem.

During the past ten years, numerical methods to solve
the relativistic or magneto-relativistic hydrodynamic
equations have advanced significantly
(see a review paper by \citet{Ibanez99} and references therein).
Especially, the method using approximate Riemann solvers
provides good accuracy even if the flow includes strong shocks
and high Lorentz factors.
Recently we have developed
a two-dimensional special relativistic code
using approximate Riemann solvers which are derived
by spectral composition of Jacobian matrices of
special relativistic hydrodynamic equations.
Plane, cylindrical($r-z$), and spherical($r-\theta$) geometry
are assumed.

Recently other new methods also have been employed to
relativistic hydrodynamic or relativistic magnetohydrodynamic codes
and proposed
(\cite{Sokolov01,DelZanna02,Annios03,DelZanna03}).

\subsection{NUMERICAL METHOD}
Relativistic hydrodynamics equations are written
in conservative form in each coordinate system,

PLANE
\begin{eqnarray}
{\partial \mbox{\boldmath $u$}\over \partial t}+
{\partial \mbox{\boldmath $f$}(\mbox{\boldmath $u$})\over \partial x}+
{\partial \mbox{\boldmath $g$}(\mbox{\boldmath $u$})\over \partial y}=0,
\end{eqnarray}

CYLINDRICAL
\begin{eqnarray}
{\partial \mbox{\boldmath $u$}\over \partial t}+
{1\over r}
{\partial (r \mbox{\boldmath $f$}(\mbox{\boldmath $u$}))
\over \partial r}+
{\partial \mbox{\boldmath $g$}(\mbox{\boldmath $u$})\over \partial z}=
\mbox{\boldmath $s_c$}(\mbox{\boldmath $u$}),
\end{eqnarray}

SPHERICAL
\begin{eqnarray}
{\partial \mbox{\boldmath $u$}\over \partial t}+
{1\over r^2}
{\partial (r^2 \mbox{\boldmath $f$}(\mbox{\boldmath $u$}))\over \partial r}+
{1\over r \sin \theta}
{\partial (\sin \theta \mbox{\boldmath $g$}(\mbox{\boldmath $u$}))
\over \partial \theta}=
\mbox{\boldmath $s_s$}(\mbox{\boldmath $u$}),
\end{eqnarray}
where, {\boldmath $u$}, {\boldmath $f$} and {\boldmath $g$}, and
{\boldmath $s_c$} and {\boldmath $s_s$} are
conservative vector, flux vectors, and source vectors,
respectively.
They are defined as follows,
\begin{eqnarray}
 \mbox{\boldmath $u$}=(\rho W,\rho h W^2 v^1,\rho h W^2 v^2,
\rho h W^2-p-\rho W)^{T},\\
\mbox{\boldmath $f$}(\mbox{\boldmath $u$})=
(\rho Wv^{1},\rho hW^2v^{1}v^{1}+p,
\rho hW^2v^{1}v^{2},
\rho hW^2v^{1}-\rho Wv^{1})^{T},\\
\mbox{\boldmath $g$}(\mbox{\boldmath $u$})=
(\rho Wv^{2},\rho hW^2v^{1}v^{2},
\rho hW^2v^{2}v^{2}+p,
\rho hW^2v^{2}-\rho Wv^{2})^{T},\\
\mbox{\boldmath $s_c$}
=\left(0,{p\over r},0,0\right)^T,\\
\mbox{\boldmath $s_s$}
=\left(0,{2p +\rho h W^2 v^2v^2 \over r},
-{\rho h W^2 v^1v^2-\cot \theta p\over r},
0\right)^T.
\end{eqnarray}

Then, we disperse these equations for the numerical calculations,
\begin{eqnarray}
\begin{array}{rcl}
\vspace{0.15cm}
\mbox{\boldmath $u$}^{n+1}_{i,j} & = 
& \mbox{\boldmath $u$}^{n}_{i,j} \\
\vspace{0.15cm}
 & - & \displaystyle{{1\over r_{i,j}}}
\left(r_{i+1/2,j}
\mbox{\boldmath $\tilde{f}$}_{i+1/2,j}-
r_{i-1/2,j}
\mbox{\boldmath $\tilde{f}$}_{i-1/2,j}\right) 
{\Delta t \over \Delta r_{i}} \\
\vspace{0.1cm}
 & - & 
(\mbox{\boldmath $\tilde{g}$}_{i,j+1/2}-
\mbox{\boldmath $\tilde{g}$}_{i,j-1/2}) 
\displaystyle{{\Delta t \over \Delta z_{j}}} \\
 & + & \mbox{\boldmath ${s}_c$}_{i,j},
\end{array}
\end{eqnarray}
where $\mbox{\boldmath $u$}^{n}_{i,j}$ stands for the average value of
$\mbox{\boldmath $u$}$ in the i,j-th grid at $t_{n}$,
and
$\mbox{\boldmath $\tilde{f}$}$ and
$\mbox{\boldmath $\tilde{g}$}$
stand for numerical flux
through the cell surface.
Here, we show the cylindrical coordinate case.
Numerical fluxes $\mbox{\boldmath $\tilde{f}$}$ and
$\mbox{\boldmath $\tilde{g}$}$
at each cell surface are calculated
by Marquina's flux formula derived from
left and right eigenvectors and eigenvalues of
Jacobian Matrices of relativistic hydrodynamic equations
\citep{Donat96,Donat98}.
To obtain higher accuracy in spatial dimensions,
we adopt the MUSCL method \citep{van77,van79} for the reconstruction
of the left and right state at each cell surface.
\begin{eqnarray}
(q_{L})_{i+1/2}=
q_{i}+{1\over 4}
\left((1-\kappa )(\bar\Delta_{-})_{i}+(1+\kappa )(\bar\Delta _{+})_{i}\right),\\
(q_{R})_{i+1/2}=
q_{i+1}-{1\over 4}
\left((1-\kappa )(\bar\Delta_{+})_{i+1}+(1+\kappa )(\bar\Delta _{-})_{i+1}\right),
\end{eqnarray}
where $q$ is the physical value for interpolation.
In our code the primitive values such as rest mass density, pressure,
and velocity are used for this reconstruction.
The accuracy of the code for spatial dimensions
is second order when a linear interpolation is adopted ($\kappa=-1$ or, $0$),
and third order when quadratic function is used for the interpolation ($\kappa=1/3$).
We use minmod limiter
to keep Total Variation Diminishing (TVD) condition
which prevents the development of numerical oscillations.
Then $\bar\Delta_{+}$ and $\bar\Delta_{-}$ are defined;
\begin{eqnarray}
(\bar{\Delta} _{+})_{i}=\mbox{minmod} (q_{i+1}-q_{i},b(q_{i}-q_{i-1})),\\
(\bar{\Delta} _{-})_{i}=\mbox{minmod} (q_{i}-q_{i-1},b(q_{i+1}-q_{i})),\\
\mbox{minmod}(a,b)\equiv
\mbox{sign}(a)\mbox{max}(0,\mbox{min}(|a|,\mbox{sign}(a)\ b)),
\end{eqnarray}
where $b$ is a parameter that satisfies
$1\le b \le (3-\kappa )/(1-\kappa )$.
The results become diffusive with small $b$.
In this study we use $\kappa =-1$ and $b=2$.
The accuracy of the time-step is first order.
Recovery of primitive values from conservative vector
\mbox{\boldmath $u$} is done by Newton-Raphson method
at every time step \citep{Aloy99a}.

\subsection{TEST CALCULATIONS}
We show two types of test calculations of 1-dimensional
relativistic hydrodynamics.
The calculation is done with a version of 2nd order for space
and 1st order for time.
400 grid points in the interested direction
are used for both cases.
As we assume the dynamics is only one dimensional,
the velocity in the other direction is set to be zero initially.
So, the velocity of this direction remains zero with time evolution.

\subsubsection{SHOCK TUBE PROBLEM}
A shock tube problem is a kind of initial value problem.
Two states are given between a discontinuity at $t=0$ initially .
This is reasonable because the analytical solution
is given by \citet{Thompson86,Marti94}.
We use two initial conditions of this problem following \citet{Donat98}.

\begin{itemize}
\item Problem 1 : SH1 (Plane)\\
Left \ \  state ($x<0.5$); $\rho_{L}=10, p_{L}=13.3, v_{L}=0, \gamma_{L}=5/3$\\
Right state ($x>0.5$); $\rho_{R}=1, p_{R}=1\times 10^{-6},
v_{R}=0, \gamma_{R}=5/3$\\

\item Problem 2 : SH2 (Plane)\\
Left \ \  state ($x<0.5$); $\rho_{L}=1, p_{L}=1000, v_{L}=0, \gamma_{L}=5/3$\\
Right state ($x>0.5$); $\rho_{R}=1, p_{R}=0.01,
v_{R}=0, \gamma_{R}=5/3$\\
\end{itemize}

\begin{figure}
\begin{center}
\resizebox{10.0cm}{!}{\plotone{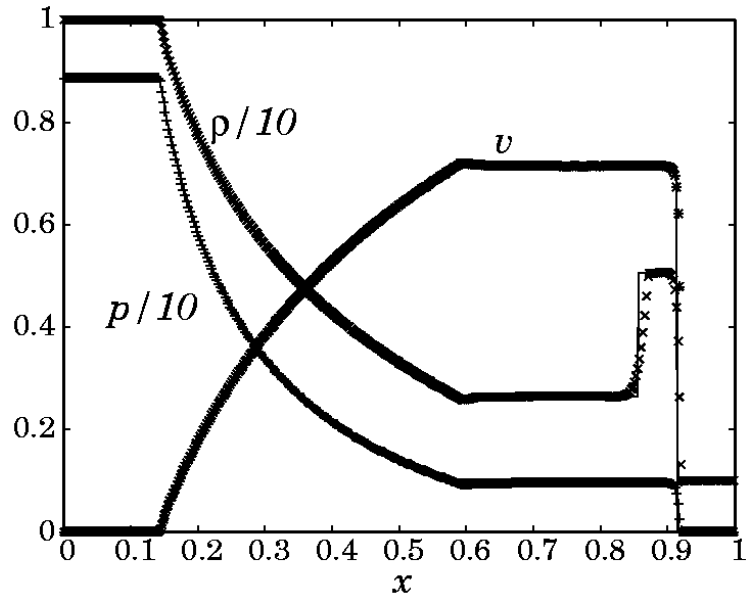}}
\caption{Test calculation of shock tube problem (SH1),
rest mass density ($\rho$), pressure ($p$),
velocity ($v$) are shown at t=0.5. Initial discontinuity is at $x=0.5$.
Solid lines are analytical solutions.
\label{SH1}}
\end{center}
\end{figure}

\begin{figure}
\begin{center}\resizebox{10.0cm}{!}{\plotone{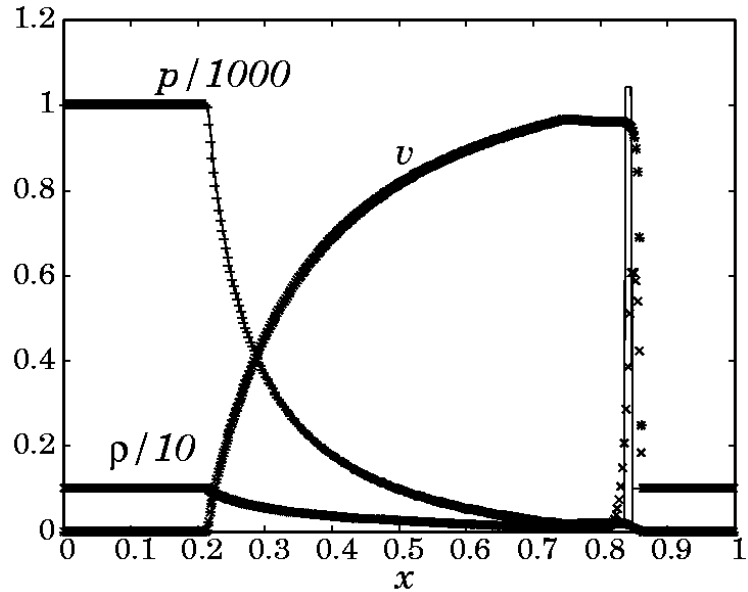}}
\caption{Same as Fig. \ref{SH1} of SH2 at $t=0.35$.\label{SH2}}
\end{center}
\end{figure}

Figures \ref{SH1} and \ref{SH2} show the results of these problems
at $t=0.5$ (SH1) and $t=0.35$ (SH2).
Analytical solutions are also shown for the comparison.
Despite of more than 5 orders jump in pressure at the initial
discontinuity,
a rarefaction fan whose front proceeds towards the left
with a sound velocity of the left side
state of initial condition, a contact discontinuity and a shock
are given with good accuracy in each case
except at the density jump behind the shock in SH2.

\subsubsection{REFLECTION SHOCK PROBLEM}
A reflection shock problem is suitable for studying a
flow with a strong shock.
Initially a relativistic homogeneous cold flow
($\rho_0,W_{0},\epsilon_0\sim0$)
reflects at $x=0$;
a wall(plane geometry), symmetric axis (cylindrical geometry)
or symmetric point (spherical geometry).
Then, a strong shock runs against the flow with a high density and
pressure jump.
Analytical solutions of this problem are given
by Rankine-Hugoniot relation of relativistic hydrodynamics
\citep{Johnson71}.
\begin{eqnarray}
\rho=\left({\gamma+1\over \gamma-1}+{\gamma\over \gamma-1}(W_0-1)\right)\rho_0,
\epsilon =W_0-1, v=0,\ \mbox{for}\  x< V_s t\\
\rho=\rho_0\left(1+{|v_0|t\over x}\right)^{\alpha},\epsilon \sim 0,v=v_0,
\ \mbox{for} \ x>V_{s}t,
\end{eqnarray}
where $V_s$ is shock velocity,
\begin{eqnarray}
V_{s}=\left({W_0-1\over W_0+1}\right)^{1/2}(\gamma-1).
\end{eqnarray}
The geometry is plane for ($\alpha =0$), cylindrical ($\alpha =1$),
and spherical ($\alpha =2$).
The expression of rest mass density jump at the shock front is
divided in two parts, namely
a maximum density compression ratio in non relativistic hydrodynamics
($(\gamma+1)/(\gamma-1)$) and including a Lorentz factor.
\begin{itemize}
\item Problem 3 : REP\\
$\rho_0=1.0,\epsilon_0=10^{-4},v_0=-0.999(W_0=22), \gamma =4/3$, Plane

\item Problem 4 : REC\\
$\rho_0=1.0,\epsilon_0=10^{-4},v_0=-0.999(W_0=22), \gamma =4/3$, Cylindrical

\item Problem 5 : RES\\
$\rho_0=1.0,\epsilon_0=10^{-4},v_0=-0.999(W_0=22), \gamma =4/3$, Spherical
\end{itemize}

\begin{figure}
\begin{center}
\resizebox{10.0cm}{!}{\plotone{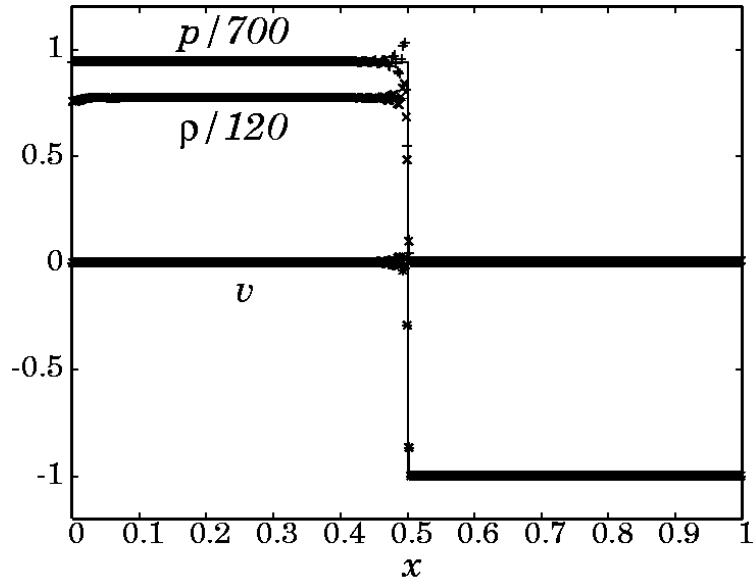}}
\caption{Test calculation of reflection shock problem (REP;plane),
rest mass density ($\rho$), pressure ($p$),
velocity ($v$) are shown at t=1.57.
Solid lines are analytical solutions.
\label{REP}}
\end{center}
\end{figure}

\begin{figure}
\begin{center}\resizebox{10.0cm}{!}{\plotone{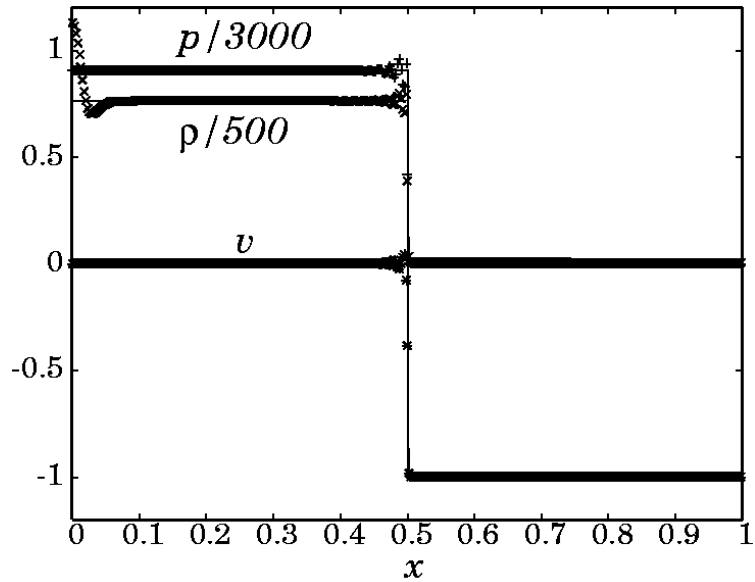}}
\caption{Same as Fig. \ref{REP} of REC(cylindrical) at $t=1.57$.\label{REC}}
\end{center}
\end{figure}

\begin{figure}
\begin{center}\resizebox{10.0cm}{!}{\plotone{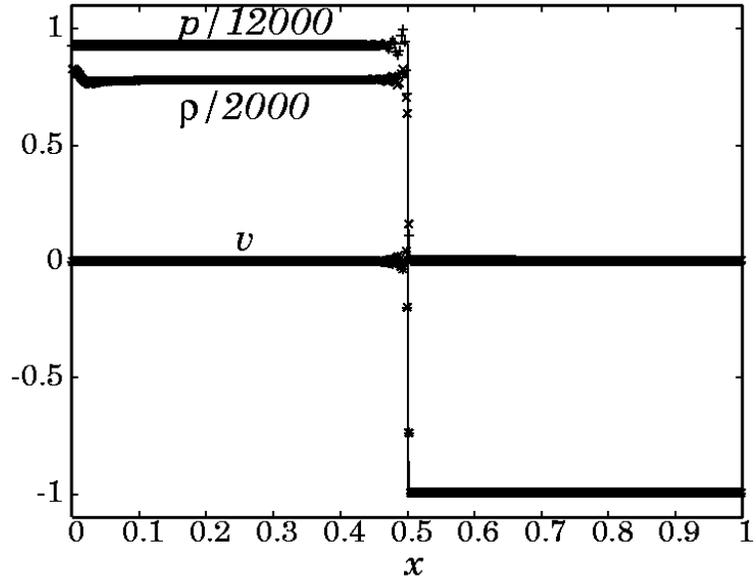}}
\caption{Same as Fig. \ref{REP} of RES(spherical) at $t=1.57$.\label{RES}}
\end{center}
\end{figure}

Figures \ref{REP}, \ref{REC}, and \ref{RES} show the numerical results
for rest mass density, pressure, and velocity for each of the problems.
Analytical solutions are also given.
We show the results at $t=1.57$ when the shock propagates 0.5 from $x=0$.
In all cases the shock front is captured with good accuracy.
Because of numerical problem,
a oscillation appears at the front.
At the boundary, the error is within 2 \% (REP), 6 \% (REC),
and 5 \% (RES) in density.
The geometry error in cylindrical and spherical case
is 1 \% (REC), and 2 \% (RES) in density.

\end{appendix}

\begin{table}[h]
\begin{center}
\begin{tabular}{c c c c}
\hline\hline
 & JB02 & JB03 & JB04 \\  \hline
 $\eta\equiv {\rho_{b}/\rho_{a}}$ 
 & $1.28\times10^{-3}$ &
$3.76\times10^{-3}$ & $9.15\times10^{-3}$\\
 $M_{b}\equiv {v_{b}/c_{b}}$& 6.0 & 6.0 & 6.0\\
 $\epsilon_{b}$&$2.55\times10^{-2}$ &
 $2.55\times10^{-2}$ & $2.55\times10^{-2}$ \\
 $\gamma$& 5/3 & 5/3 & 5/3 \\
 $K\equiv p_{b}/p_{a}$ & 10 & 33 & 100\\
 $W_{b}(v_{b})$ & 7.1(0.99) & 7.1(0.99) & 7.1(0.99)\\
 $v_{j}^{1D}$ & 0.2 & 0.3 & 0.4\\
 expiration time $\mbox{R}_{\mbox{b}}/(c)$& 1800 & 1200 & 600\\
                                \hline
\end{tabular}
\caption{numerical conditions of models}\label{condition}

\end{center}
\end{table}

\clearpage

\begin{figure}
\caption{Contours of rest mass density(top), pressure(middle), and
Lorentz factor(bottom) of model JB02 at the end of the simulation
 ($t=1770[\mbox{R}_{b}/c]$)
\label{contourA}}
\end{figure}
\clearpage

\begin{figure}
\caption{Contours of rest mass density(top), pressure(middle), and
Lorentz factor(bottom) of model JB03 at the end of the simulation
 ($t=1140[\mbox{R}_{b}/c]$)
\label{contourB}}
\end{figure}

\clearpage

\begin{figure}
\caption{Contours of rest mass density(top), pressure(middle), and
Lorentz factor(bottom) of model JB03 at the end of the simulation
 ($t=570[\mbox{R}_{b}/c]$)
\label{contourC}}
\end{figure}

\clearpage

\clearpage

\begin{figure}
\caption{The absolute velocity ($r>0$) and log scale rest mass density contour ($r<0$) of
model JB02 (top), JB03 (middle), and JB04 (bottom) in early phase,
at $t=107.5\mbox{R}_{\mbox{b}}/c$(JB02),
$t=50.0\mbox{R}_{\mbox{b}}/c$(JB03),
$t=35.0\mbox{R}_{\mbox{b}}/c$(JB04)
\label{velocity}}
\end{figure}

\clearpage

\clearpage

\clearpage

\begin{figure}
\caption{A series of profiles of
growing and separating of vortices.
In each panel the top shows the
absolute velocity $r>0$ and the bottom shows rest mass density $r<0$,
(a):normal profile, (b):an oblique shock appears,
(c)-(e):a vortex grows up a vortex,
(f)-(g):separates from top of the jet
\label{vortex}}
\end{figure}

\clearpage

\begin{figure}
\begin{center}
\caption{Synchrotron emissivity of JB02 at $t=1770[\mbox{R}_{b}/c]$
with several angle $\theta=0, 15, 30, 45, 60, 75,$ and $90 ^{\circ}$,
where $\theta$ is the angle between $z$ axis and the line of sight.
Counters are shown in log scale with four order magnitude from maximum intensity
using gray scaled color bar (white is the highest emissivity)
in each panel.
  \label{emissionJB02}}
\end{center}
\end{figure}

\clearpage

\begin{figure}
\begin{center}
\caption{Same as Fig. \ref{emissionJB02} of JB04 at
 $t=570[\mbox{R}_{b}/c]$.
\label{emissionJB03}}
\end{center}
\end{figure}

\clearpage

\clearpage

\clearpage

\clearpage

\end{document}